\documentclass[a4paper,fleqn,usenatbib, useAMS]{mnras}

\usepackage{newtxtext,newtxmath}

\usepackage[T1]{fontenc}
\usepackage{ae,aecompl}
\usepackage{graphicx}	
\usepackage{amsmath}	
\usepackage{amssymb}	
\usepackage{bm}

\def\be{\begin{equation}}
\def\ee{\end{equation}}
\def\bea{\begin{eqnarray}}
\def\eea{\end{eqnarray}}

\def\ba#1\ea{\begin{align}#1\end{align}}
\def\bg#1\eg{\begin{gather}#1\end{gather}}

\renewcommand{\v}[1]{\bm{#1}}

\newcommand{\vx}{\v{x}}

\newcommand{\vk}{\v{k}}

\newcommand{\vtheta}{\v{\theta}}
\newcommand{\vell}{\v{\ell}}

\def\be{\begin{equation}}
\def\ee{\end{equation}}
\def\ben{\begin{eqnarray}}
\def\een{\end{eqnarray}}
\def\ba{\begin{array}}
\def\ea{\end{array}}

\def\ba#1\ea{\begin{align}#1\end{align}}

\newcommand{\bq}{\begin{eqnarray}}
\newcommand{\eq}{\end{eqnarray}}
\newcommand{\bes}{\begin{subequations}}
\newcommand{\ees}{\end{subequations}}

\def\tt{t_{\tilde 0}}

\newcommand{\fidu}{Fiducial}
\newcommand{\high}{SepHigh}
\newcommand{\loww}{SepLow}

\def\P{\mathcal{P}}

\def\O{\mathcal{O}}

\def\S{\mathcal{S}}

\makeatletter
\newlength{\apb@width}
\newcommand{\autoparbox}[2][c]{\settowidth{\apb@width}{#2}\parbox[#1]{\apb@width}{#2}}

\makeatother

\DeclareMathOperator{\cov}{Cov}

\newcommand{\comment}[1]{}

\title[Separate Universe TNG]{Separate Universe Simulations with IllustrisTNG: baryonic effects on power spectrum responses and higher-order statistics}

\author[Barreira et al.]{Alexandre Barreira,$^{1}$\thanks{E-mail: barreira@mpa-garching.mpg.de}
Dylan Nelson,$^{1}$
Annalisa Pillepich,$^{2}$
Volker Springel,$^{1}$
\newauthor
Fabian Schmidt,$^{1}$
Ruediger Pakmor,$^{1}$
Lars Hernquist,$^{3}$
Mark Vogelsberger$^{4}$
\\
$^{1}$Max-Planck-Institut f\"ur Astrophysik, Karl-Schwarzschild-Str. 1, 85741 Garching, Germany\\
$^{2}$Max-Planck-Institut f\"ur Astronomie, K\"onigstuhl 17, 69117 Heidelberg, Germany \\
$^{3}$Harvard-Smithsonian Center for Astrophysics, 60 Garden Street, Cambridge, MA 02138 \\
$^{4}$Department of Physics, Kavli Institute for Astrophysics and Space Research, MIT, Cambridge, MA 02139, USA
}

\begin{document}

\label{firstpage}
\pagerange{\pageref{firstpage}--\pageref{lastpage}}
\maketitle

\begin{abstract} 
We measure power spectrum response functions in the presence of baryonic physical processes using separate universe simulations with the IllustrisTNG galaxy formation model. The response functions describe how the small-scale power spectrum reacts to long-wavelength perturbations and they can be efficiently measured with the separate universe technique by absorbing the effects of the long modes into a modified cosmology. Specifically, we focus on the total first-order matter power spectrum response to an isotropic density fluctuation $R_1(k,z)$, which is fully determined by the logarithmic derivative of the nonlinear matter power spectrum ${\rm dln}P_m(k,z)/{\rm dln}k$ and the growth-only response function $G_1(k,z)$. We find that $G_1(k,z)$ is not affected by the baryonic physical processes in the simulations at redshifts $z < 3$ and on all scales probed ($k \lesssim 15h/{\rm Mpc}$, i.e. length scales $\gtrsim 0.4 {\rm Mpc}/h$). In practice, this implies that the power spectrum fully specifies the baryonic dependence of its response function. Assuming an idealized lensing survey setup, we evaluate numerically the baryonic impact on the squeezed-lensing bispectrum and the lensing super-sample power spectrum covariance, which are given in terms of responses. Our results show that these higher-order lensing statistics can display varying levels of sensitivity to baryonic effects compared to the power spectrum, with the squeezed-bispectrum being the least sensitive. We also show that ignoring baryonic effects on lensing covariances slightly overestimates the error budget (and is therefore conservative from the point of view of parameter error bars) and likely has negligible impact on parameter biases in inference analyses.
\end{abstract}


\section{Introduction}\label{sec:intro}

The large-scale distribution of matter in the Universe is one of the best tools to address some of the deepest open problems in fundamental physics. These include studies of the nature of dark energy and dark matter, tests of the law of gravity on large scales, tests of early-Universe inflationary models, as well as constraints on the absolute value of neutrino masses. Within large-scale structure analyses, most attention has been devoted to studies of the clustering pattern of galaxies, which trace the underlying clustering of matter in a biased manner, and to studies of background galaxy shapes that are coherently distorted (cosmic shear) by intervening gravitational potentials via weak gravitational lensing. Observational analyses of galaxy clustering \citep{2017MNRAS.470.2617A, 2017MNRAS.466.2242B, 2017MNRAS.464.1640S, 2017MNRAS.465.1757G} and weak lensing data \citep{2017MNRAS.465.1454H, 2018PhRvD..98d3526A, 2018MNRAS.474.4894J, 2018arXiv180909148H} have provided tighter and tighter constraints on various cosmological parameters over the years, and are expected to continue to do so as next-generation surveys like DESI \citep{2013arXiv1308.0847L}, LSST \citep{DESC-SRD} and Euclid \citep{2011arXiv1110.3193L} come online.

The most popular way to characterize the statistical information encoded in large-scale structure is via the two-point correlation function, or its Fourier transform, the power spectrum. In galaxy clustering studies, the main theoretical modeling steps involve descriptions of galaxy bias \citep{biasreview} and redshift-space distortions. The most frequently used methods are based on perturbation theory \citep{Bernardeau/etal:2002}, whose limited accuracy in the nonlinear regime of structure formation has been restricting most analyses to sufficiently large-scales $k \lesssim 0.3\ h/{\rm Mpc}$ ($\gtrsim 30\ {\rm Mpc}$). Weak-lensing analyses, on the other hand, are sensitive to the total clustering of matter, which is better understood in the nonlinear regime and permits extending the analysis to smaller length scales. Here, an important factor in setting the smallest length scale used in observational analyses is associated with the uncertain impact that baryonic physical processes can have on the small-scale power spectrum. A number of different hydrodynamical simulations \citep{2011MNRAS.415.3649V, 2016MNRAS.461L..11H, 2018MNRAS.475..676S, 2018MNRAS.480.3962C} have consistently reported a suppression of the total clustering power on scales $k \gtrsim 1 h / {\rm Mpc}$, but the precise shape and magnitude of the effect depend on the exact implementation of the baryonic physical processes, most notably feedback effects by active galactic nuclei (AGN) and the corresponding impact on halo dark matter and gas profiles \citep{2018arXiv181008629S}. Current weak lensing analyses circumvent this problem by either simply discarding the data in regimes that may be affected by baryons, or modeling the impact of baryons with a number of nuisance parameters that are subsequently marginalized over. Given that a significant part of the constraining power in weak-lensing surveys comes from small scales, the development of strategies to incorporate or mitigate the impact of baryonic effects is a pressing issue to address. This has consequently become the subject of recent active work on lensing power spectrum analyses \citep{Semboloni:2011fe, 2015JCAP...12..049S, 2015MNRAS.454.2451E, 2015MNRAS.454.1958M, 2015MNRAS.450.1212H, 2018MNRAS.480.3962C, 2018arXiv180901146H, 2018arXiv181008629S}.

The fact that the late-time distribution of matter in the Universe is appreciably non-Gaussian distributed immediately leads, however, to the observation that the power spectrum (despite being the most popular {\it summary statistic}) cannot describe all of the available information. This naturally motivates incorporating higher-order $N$-point correlation functions in observational analyses as a way to gain access to the information that is missed by the power spectrum, and consequently, obtain improved constraints on cosmological parameters. For instance, the gains that can be achieved by incorporating the bispectrum (the Fourier transform of the three-point function) in cosmological analyses are by now well established \citep{2001ApJ...548....7C, 2005PhRvD..72h3001D, 2006PhRvD..74b3522S, 2013PhRvD..87l3538S, 2013arXiv1306.4684K, 2017MNRAS.465.1757G, 2017PhRvD..96b3528C, 2018arXiv181002374C, 2018arXiv181207437R, 2018MNRAS.tmp.2989Y}. The body of work on higher-order $N$-point functions in the nonlinear regime is however much smaller compared to that on the power spectrum. The reason for this can be traced back to the fact that the estimators of the $N$-point functions become rapidly numerically intensive with increasing $N$ \citep{2019arXiv190100296S}, as well as the fact that these $N$-point functions live in higher-dimensional parameter spaces that require large-volume, high-resolution cosmological simulations to obtain sufficiently precise measurements. Note also that knowledge of higher-order statistics is important even for power spectrum analysis since its covariance is given by a specific configuration of the four-point correlation function \citep{1999ApJ...527....1S}.

The larger numerical resources needed to study higher-order correlation functions become even more critical for the case of hydrodynamical simulations, which are appreciably more demanding to run than equivalent gravity-only counterparts. This helps explain the relatively small number of results on the impact of baryonic physics on higher-order lensing correlation functions, which remains a largely unexplored subject in the literature, despite a few interesting existing studies. For instance, \cite{2013MNRAS.434..148S} used results from the OWLS suite of simulations  \citep{2010MNRAS.402.1536S} to predict the impact of baryonic physics on the three-point lensing correlation function. Further, \cite{2015ApJ...806..186O} investigated the impact of baryonic effects on weak-lensing peaks and Minkowski functionals, which are statistics sensitive to higher-order $N$-point functions (see also \cite{2013PhRvD..87b3511Y}, and \cite{2018MNRAS.478.1305C} for a study of baryonic effects on the probability distribution function of weak-lensing maps). Studies such as these can provide interesting insights on the physics of nonlinear structure formation and are therefore crucial to fully exploit the constraining power of existing and future datasets.

In this paper, we take a number of steps forward in understanding the impact of baryonic effects on higher-order $N$-point correlation functions by using the response approach to perturbation theory developed by \cite{responses1, responses2} as an extension of previous work done by \cite{2014JCAP...05..048C, response, 2015JCAP...09..028C}. Power spectrum responses are functions of scale and time that encode how the small-scale power spectrum {\it responds} (or reacts) to the presence of long-wavelength density and tidal field perturbations. These response functions can be measured efficiently in the nonlinear regime using so-called separate universe simulations \citep{li/hu/takada, 2014PhRvD..90j3530L, wagner/etal:2014, CFCpaper2, baldauf/etal:2015, response, andreas}, which are simulations that absorb the impact of the long-wavelength modes into a redefinition of the cosmology that is simulated. In a perturbation theory sense, these response functions describe the coupling between small- and large-scale Fourier modes, quite importantly, for nonlinear values of the small-scale modes. In practice, this permits one to straightforwardly evaluate a number of higher-order correlation functions in the nonlinear regime \citep{responses1}. Importantly, accurate and precise measurements of the response functions can be obtained with the same particle resolution and simulation volume as for the matter power spectrum, thereby permitting systematic studies of higher-order correlation functions without exacerbated demand for numerical resources.

Concretely, in this paper we apply the separate universe simulation technique using the IllustrisTNG galaxy formation model \citep{2017MNRAS.465.3291W, Pillepich:2017jle} to measure for the first time the matter power spectrum responses in the presence of baryonic physical processes. All past matter power spectrum response measurements \citep{li/hu/takada, response, baldauf/etal:2015, andreas} have been carried out using gravity-only separate universe simulations. Once the dependence of the power spectrum and its response functions on baryonic effects is determined, the machinery of the response approach to perturbation theory \citep{responses1} can then be used to predict the corresponding impact of baryons on a number of higher-order correlation functions. We shall do so explicitly for the case of the squeezed weak-lensing convergence bispectrum, as well as for the super-sample covariance (SSC; \cite{takada/hu:2013, completessc}) part of the covariance matrix of the weak lensing convergence power spectrum (which is a four-point function). 

The rest of this paper is organized as follows. In Sec.~\ref{sec:theory}, we outline the main concepts behind matter power spectrum responses and the separate universe simulations we performed. In Sec.~\ref{sec:results1}, we present and discuss our numerical measurements of the power spectrum response functions with the IllustrisTNG model. In Sec.~\ref{sec:results2}, we use the response approach to perturbation theory to calculate the impact of baryonic effects on the squeezed lensing bispectrum and lensing power spectrum covariances. We summarize and conclude in Sec.~\ref{sec:sum}.

\section{Power spectrum responses and Separate Universe simulations}\label{sec:theory}

In this section, we outline the main concepts behind power spectrum response functions and separate universe simulations. These have been discussed at length in past works and we thus refer the interested reader to the cited literature for more details.

\subsection{Power spectrum responses}\label{sec:theory_R}

Power spectrum responses are functions that describe how the local, nonlinear matter power spectrum responds to the presence of long-wavelength perturbations. Formally, we can write \citep{responses1}
\bq\label{eq:Pk_exp}
P_m(\vk, t | \O(\vx, t)) = P_m(k,t) \left[1 + \sum_{\O} R_\O(k,t) \O(\vx,t)\right],
\eq
where $P_m(\vk, t | \O(\vx,t))$ is the nonlinear matter power spectrum at physical time $t$ measured in some local volume around position $\vx$ that is embedded in a long-wavelength perturbation $\O$, $P_m(k,t)$ is the global matter power spectrum measured at cosmic mean density and $R_\O(k,t)$ are response coefficient functions that specify the response of the power spectrum to each specific long-wavelength perturbation $\O$ (see also \cite{bertolini1}). The meaning of the {\it local} power spectrum is that it is measured in a region of size $L$ that is sufficiently smaller than the wavelength of the perturbation $\O$, i.e., $1/k < L \ll 1/p$, where $p$ is the Fourier mode of the perturbation $\O$.

The total matter power spectrum response is specified by listing all possible long-wavelength perturbations $\O$ the local power spectrum can depend on. The leading order ones (see \cite{responses1} for a discussion) are those that involve second spatial derivatives of the gravitational potential $\Phi$, such as, isotropic density perturbations $\O(\vx, t) = \delta(\vx, t) \sim \nabla^2 \Phi(\vx, t)$ or tidal field perturbations $\O(\vx,t) = \hat{k}^i\hat{k}^jK_{ij}(\vx,t) = \hat{k}^i\hat{k}^j \left[\nabla_i\nabla_j/\nabla^2 - \delta_{ij}/3\right]\delta(\vx,t)$ (note how for tidal fields there is an explicit dependence on the direction of the small-scale mode $\vk$, which is why the  local power spectrum on the left-hand side of Eq.~(\ref{eq:Pk_exp}) can be anisotropic in general \citep{akitsu/takada/li, 2018PhRvD..97f3527A, 2018JCAP...02..022L, andreas}). In this paper, we focus on isotropic density fluctuations, for which Eq.~(\ref{eq:Pk_exp}) specializes to
\bq\label{eq:Pk_expiso}
P_m(\vk, t | \delta_L(\vx, t)) &=& P_m(k,t) \left[1 + \sum_{n=1}^N \frac{R_n(k,t)}{n!}\delta_L(\vx, t)^n\right]
\eq
i.e., the response functions $R_n(k,t)$ can be interpreted as the coefficients of a Taylor expansion of the local power spectrum in the long-wavelength isotropic overdensity $\delta_L(\vx, t)$ (the subscript $_L$ emphasizes the long-wavelength nature of the perturbation). For small enough $\delta_L$ values, we can linearize Eq.~(\ref{eq:Pk_expiso}) and solve for the first-order isotropic power spectrum response
\bq\label{eq:R1}
R_1(k, t) = \frac{1}{\delta_L(\vx, t)}\left[\frac{P_m(\vk, t | \delta_L(\vx, t))}{P_m(k,t)} - 1\right],
\eq
which is the response function we wish to measure below with separate universe simulations of the IllustrisTNG model (see \cite{response} for measurements of higher-order isotropic responses, $R_n$ $(n>1)$, in gravity-only simulations).

The presence of the long-wavelength perturbation $\delta_L(x,t)$ induces three physically distinct effects on the small-scale power spectrum \citep{li/hu/takada, response}:

\begin{enumerate}
\item {\it Reference density effect.} This accounts for the fact that measurements of the local power spectrum are performed with respect to a modified mean density $\rho_m(\vx,t) = \rho_m(t)\left[1 + \delta_L(\vx, t)\right]$, where $\rho_m(t)$ is the cosmic mean density w.r.t.~which the global power spectrum is defined \footnote{We distinguish background-averaged quantities by omitting their dependence on the spatial coordinate $\vx$.}.

\item {\it Dilation effect.} This describes the rescaling of the local scale factor $a(\vx, t)$, which evolves differently than the global one $a(t)$ because of the long-wavelength mode. In other words, the expansion of spacetime inside the perturbation $\delta_L(\vx, t)$ takes place at a different rate than the global expansion of the Universe. In practice, this induces a rescaling of the physical scales inside the perturbation.

\item {\it Growth-only effect.} This describes the actual mode-coupling between the linear, long-wavelength mode and nonlinear small-scale power spectrum modes.
\end{enumerate}

Accounting for these effects permits one to write the first-order response $R_1(k, t)$ as
\bq\label{eq:R1_exp}
R_1(k, t) = 1 - \frac{1}{3} \frac{{\rm dln}P_m(k,t)}{{\rm dln}k} + G_1(k,t),
\eq
where $G_1(k,t)$ is called the growth-only response function that describes effect (iii) above in isolation. Strictly speaking, this is the only term for which separate universe simulations are needed, with the remainder of the response function $R_1$ being completely specified by the logarithmic derivative of the nonlinear matter power spectrum ${\rm dln}P_m(k,t)/{\rm dln}k$.

In some of our results below in Sec.~\ref{sec:results2}, we will also use the first-order response to a long-wavelength tidal field, which admits the following decomposition
\bq\label{eq:RK_exp}
R_K(k,t) = G_K(k, t) - \frac{{\rm dln}P_m(k,t)}{{\rm dln}k},
\eq
where $G_K(k, t)$ is the growth-only response to a tidal field \citep{andreas}.

\subsection{Separate Universe formalism}\label{sec:theory_sepuni}

The following observation forms the basis of the implementation of the separate universe formalism to measure power spectrum responses {\citep{li/hu/takada, 2014PhRvD..90j3530L, wagner/etal:2014, CFCpaper2, baldauf/etal:2015, response, andreas}: {\it local structure formation in a given cosmology in the presence of a long-wavelength perturbation is equivalent to global structure formation in an appropriately modified cosmology.} For the particular case of isotropic density perturbations $\delta_L$ that we focus on in this paper, this corresponds to a modified cosmology with a modified mean background density
\bq\label{eq:rho_change}
\tilde{\rho}_m(t) = \rho_m(t)\left[1 + \delta_L(t)\right],
\eq
where, here and throughout, quantities with and without a tilde are defined in the modified and fiducial cosmologies, respectively. In Eq.~(\ref{eq:rho_change}), we have dropped the coordinate argument $\vx$ as we assume that the long-wavelength perturbation $\delta_L$ is effectively constant inside the local volume where one wishes to measure the power spectrum; this is why the effect can be mimicked by rescaling the mean matter density everywhere in the modified cosmology. Equation (\ref{eq:rho_change}) can be written as
\bq\label{eq:rho_change_2}
\frac{\tilde{\Omega}_{m0} \tilde{h}^2}{\tilde{a}(t)^3} = \frac{\Omega_{m0} h^2}{a(t)^3} \left[1 + \delta_L(t)\right] ,
\eq
where $\Omega_{m0}$ is the fractional matter density parameter and $h$ is the dimensionless Hubble constant $H_0 = 100h\ {\rm km/s/Mpc}$. Owing to the different mean background densities, the expansion of spacetime takes place at different rates in the two cosmologies, i.e., $a(t) \neq \tilde{a}(t)$. We adopt the usual convention that $a(t_0) = 1$ (where $t_0$ is the present-day physical time in the fiducial cosmology), and at some sufficiently early epoch $t_i$, when the amplitude of the long-mode can be neglected, we take $a(t_i) = \tilde{a}(t_i)$. In this early time limit, Eq.~(\ref{eq:rho_change_2}) becomes
\bq\label{eq:rho_change_3}
\tilde{\Omega}_{m0} \tilde{h}^2 = \Omega_{m0} h^2.
\eq
This equation can be plugged back into Eq.~(\ref{eq:rho_change_2}) to derive
\bq\label{eq:delta_a}
1 + \delta_L(t) = \left[1 + \delta_a(t)\right]^{-3},
\eq
where $\tilde{a}(t) = a(t)\left[1+\delta_a(t)\right]$.

In this paper, our fiducial cosmology is given by a spatially-flat $\Lambda{\rm CDM}$ model, for which the Friedmann equation is given by
\bq\label{eq:Friedmann_fidu}
H^2(t) = \left(\frac{\dot{a}(t)}{a(t)}\right)^2 = \frac{8\pi G}{3} \left[\rho_m(t) + \rho_{\Lambda}\right],
\eq
where an overdot denotes a derivative w.r.t.~physical time (we ignore the impact of radiation and massive neutrinos on the expansion rate). For the modified cosmology, the Friedmann equation follows as
\bq\label{eq:Friedmann_sepu}
\tilde{H}^2(t) = \left(\frac{\dot{\tilde{a}}(t)}{\tilde{a}(t)}\right)^2 = \frac{8\pi G}{3} \left[\tilde{\rho}_m(t) + \rho_{\Lambda}\right] - \tilde{K}\tilde{a}(t)^{-2},
\eq
where we have allowed for non-zero spatial curvature and the physical energy density of the cosmological constant $\rho_\Lambda$ is not affected by the long-wavelength density mode. The value of $\tilde{K}$ can be derived by taking the difference of Eqs.~(\ref{eq:Friedmann_fidu}) and (\ref{eq:Friedmann_sepu}) 
\bq\label{eq:K}
\tilde{K}a(t)^{-2} &=& \frac{8\pi G}{3} \rho_m(t) \left[\left[1 + \delta_a(t)\right]^{-1} -\left[1 + \delta_a(t)\right]^{2} \right] - \dot{\delta}_a^2 \nonumber \\
&& - 2H(t)\left[1 + \delta_a(t)\right]\dot{\delta}_a(t).
\eq
Given that $\tilde{K}$ is a constant, we can derive its value at any time using Eq.~(\ref{eq:K}). At early times, the expansion of the Universe is matter-dominated, $H^2(t) \approx \Omega_{m0}H_0^2a(t)^{-3}$, and we can linearize Eq.~(\ref{eq:delta_a}) to get $\delta_a(t) \approx -\delta_L(t)/3$; further noting that during matter domination $\delta_L(t) \propto a(t)$, we have $\dot{\delta}_a(t) = H(t)\delta_a(t)$. We can thus write
\bq\label{eq:K_1}
\frac{\tilde{K}}{H_0^2} = \frac{5}{3} \Omega_{m0} \frac{\delta_{L0}}{D(t_0)},
\eq
where $\delta_{L0}$ is the value of the long-wavelength overdensity today, and $D(t)$ is the linear growth factor of the fiducial cosmology that obeys
\bq\label{eq:lingrowth}
\ddot{D}(t) + 2H(t)\dot{D}(t) - 4\pi G \rho_m(t)D(t) = 0.
\eq
For the definition of cosmological parameters such as $H_0$ and $\Omega_{m0}$, $N$-body/hydro codes define their numerical values at the time the scale factor is equal to unity, but in the modified and fiducial cosmologies this occurs at different physical times. The parameters that are used in the code should thus be evaluated at a time $\tilde{t}_0$ defined as $\tilde{a}(\tilde{t}_0) = 1$:
\bq
\label{eq:paramdefs_1}\tilde{H}_0 &\equiv& \tilde{H}(\tilde{t}_0) \equiv H_0\left[1 + \delta_H\right]\\
\label{eq:paramdefs_2}\tilde{\Omega}_{m0} &\equiv& \frac{8\pi G}{3 \tilde{H}_0^2} \tilde{\rho}_m(\tilde{t}_0) = \Omega_{m0} \left[1 + \delta_H\right]^{-2} \\
\label{eq:paramdefs_3}\tilde{\Omega}_{\Lambda0} &\equiv& \frac{8\pi G}{3 \tilde{H}_0^2} \rho_\Lambda = \Omega_{\Lambda0} \left[1 + \delta_H\right]^{-2},
\eq
where the quantity $\delta_H$ is defined by Eq.~(\ref{eq:paramdefs_1}) and is used subsequently in Eqs.~(\ref{eq:paramdefs_2}) and (\ref{eq:paramdefs_3}). The fractional curvature density is given by $\tilde{\Omega}_{K0} = -\tilde{K}/\tilde{H}_0^2$. What is left to specify is the value of $\delta_H$, which can be obtained from the present-day value of the Friedmann equation in the modified cosmology:
\bq\label{eq:delta_H}
1 &=& \tilde{\Omega}_{m0} + \tilde{\Omega}_{\Lambda0} + \tilde{\Omega}_{K0} \nonumber \\ 
1 &=& \left(\Omega_{m0} + \Omega_{\Lambda0}\right)\left[1 + \delta_H\right]^{-2} - \tilde{K}/\tilde{H}_0^2 \nonumber \\
&\Longrightarrow& \delta_H = \sqrt{1 - \frac{\tilde{K}}{H_0^2}} - 1,
\eq
where we have used that the fiducial cosmology is spatially flat $\Omega_{m0} + \Omega_{\Lambda0} = 1$. This finalizes the derivation of the cosmological parameters of the modified cosmology given the parameters of the fiducial cosmology, as well as the present-day value of the long-wavelength mode $\delta_{L0}$.

\subsection{Separate Universe simulations}\label{sec:sims}

\begin{table}
\centering
\begin{tabular}{@{}lccccccccccc}
\hline\hline
Name &\ \  $\Omega_{m0}$ &  \ \ $\Omega_{b0}$ & \ \ $\Omega_{\Lambda 0}$ & \ \ $h$ & \ \ $L_{\rm box}\ [\frac{{\rm Mpc}}{h}]$
\\
\hline
\fidu &\ \  $0.3089$ & \ \ $0.0486$ & \ \ $0.6911$ & \ \ $0.6774$ & \ \ $205$ 
\\
\\
\high &\ \  $0.3194$ & \ \ $0.0502$ & \ \ $0.7146$ & \ \ $0.6662$ & \ \ $201.608$
\\
\\
\loww &\ \  $0.2991$ & \ \ $0.0471$ & \ \ $0.6691$ & \ \ $0.6884$ & \ \ $208.337$
\\
\hline
\hline
\end{tabular}
\caption{Parameters of the cosmologies simulated in this paper. The \high\  and \loww\ cosmologies mimic, respectively, the effect of $\delta_{L0} = 0.05$ and $\delta_{L0} = -0.05$ long-wavelength density perturbations in the \fidu\ cosmology. The comoving box size used is quoted in units of ${\rm Mpc}/h$ in the corresponding cosmology. The value of the spectral index $n_s = 0.967$ is the same for the three cosmologies, which share also the same amplitude of the initial power spectrum: that which corresponds to $\sigma_8(z=0) = 0.816$ in the \fidu\ cosmology. We have performed simulations with two particle/mass element numbers $N_p = 625^3$ (called TNG300-3), $N_p = 1250^3$ (called TNG300-2) and with (dubbed Hydro) and without (dubbed Gravity) hydrodynamical processes (for the Hydro runs, the particle number is actually twice the quoted values: $N_p$ gas cells and $N_p$ dark matter mass elements). The same random seed was used to generate the initial conditions of all the simulations. The original IllustrisTNG simulations comprise also a higher-resolution set called TNG300-1 ($N_p = 2500^3$), whose spectra \citep{2018MNRAS.475..676S} can be used to draw considerations on the convergence of our TNG300-2 results.}
\label{table:params}
\end{table}

The simulation results we present in this paper were obtained using the moving-mesh code {\sc AREPO} \citep{2010MNRAS.401..791S, 2016MNRAS.455.1134P} with the IllustrisTNG hydrodynamical galaxy formation model
\citep{2017MNRAS.465.3291W, Pillepich:2017jle}, which is an upgraded version of the original Illustris model \citep{2014MNRAS.445..175G, 2014MNRAS.444.1518V}. In this paper, we focus on the impact of baryons on matter power spectrum responses, and hence, our results are complementary to those of \cite{2018MNRAS.475..676S} who investigated already the corresponding impact on the power spectrum itself, for the same implementation of the TNG model. Here, for brevity, we skip describing the details of the IllustrisTNG model and refer the reader to the cited literature for more details, including \cite{2018MNRAS.480.5113M, Pillepich:2017fcc, 2018MNRAS.477.1206N, 2018MNRAS.475..676S, Nelson:2017cxy} for the first results from the IllustrisTNG simulations and \cite{Nelson:2018uso} for an overview of the publicly available simulation data.

The cosmological parameters of the fiducial and separate universe simulations we consider are summarized in Table \ref{table:params}. The \fidu\ cosmology parameters are the same as in the main IllustrisTNG runs;  the two separate universe runs \high\ and \loww\ correspond to $\delta_{L0} = 0.05$ and $\delta_{L0} = -0.05$, respectively. This choice for the amplitude of the long-wavelength modes is motivated by having a strong enough perturbation capable of inducing sizeable changes to the power spectrum, but keeping higher order corrections $\mathcal{O}(\delta_L^2)$ in Eq.~(\ref{eq:Pk_expiso}) negligible.  Within this linear regime ($\delta_{L0} \ll 1$), any two values of $\delta_{L0}$ are sufficient to make one measurement of the first-order response by finite differencing the corresponding power spectra. Here, we opt to utilize three values values ($\delta_{L0} = 0, \pm 0.05$) to be able to make a measurement using an overdensity and an underdensity (note however that the result should be independent of the sign of $\delta_{L0}$); cf.~Eqs.~(\ref{eq:G1measure})-(\ref{eq:G1measure3}) below). The simulation box size of the Fiducial cosmology is $L_{\rm box} = 205 {\rm Mpc}/h \approx 300 {\rm Mpc}$ and we consider two particle/mass element resolutions: $N_p = 625^3$ and $N_p = 1250^3$, to which we refer below as TNG300-3 and TNG300-2, respectively.\footnote{This is the same terminology as that used in the main IllustrisTNG runs. In the latter, for this box size, there is a higher-resolution set called TNG300-1 with $N_p = 2500^3$. In this paper, we are interested in the clustering of matter on scales $k \lesssim 10\ h/{\rm Mpc}$, on which \cite{2018MNRAS.475..676S} has shown that the TNG300-1 and TNG300-2 runs display $\%$-level agreement.} To isolate the impact of baryonic physics on the power spectrum responses we have run standard IllustrisTNG hydrodynamical simulations (dubbed Hydro below), as well as gravity-only simulations without baryonic processes (dubbed Gravity below).

The initial conditions were generated with the {\sc N-GenIC} code \citep{2015ascl.soft02003S} using the Zel'dovich approximation at the same initial redshift $z_i = 127$ for all the cosmologies (this ignores the fact that $z(t) \neq \tilde{z}(t)$, but which is a small effect at early times).\footnote{The use of the Zel'dovich approximation, although not as accurate as second-order perturbation theory methods (2LPT), is not expected to have a visible impact on our results since the simulations start at sufficiently high redshift and our main results consist of power spectra ratios, and not their absolute values.} The shape of the linear matter power spectrum is the same in the three cosmologies because the physical matter densities are the same (cf.~Eq.~(\ref{eq:rho_change_3})). We generate the initial power spectrum file using the {\sc CAMB} code \citep{camb, 2011ascl.soft02026L} at $z=0$ and rescale it to $z_i$ using the growth factor $D(t)$ of the fiducial cosmology. An extra step that is needed for the modified cosmologies is to change the units of the wavenumbers and power spectrum from $h/{\rm Mpc}$ and ${\rm Mpc}^3/h^3$ to $\tilde{h}/{\rm Mpc}$ and ${\rm Mpc}^3/\tilde{h}^3$; effectively, this is done by multiplying the $k$ values by $\left(1+\delta_H\right)^{-1}$ and the power spectra values by $\left(1+\delta_H\right)^{3}$.

Another noteworthy consideration concerns the choice of the box size for the \high\ and \loww\ simulations. Given that the measurement of the responses is effectively the ratio of the spectra in the modified and fiducial cosmologies (cf.~Eq.~(\ref{eq:R1})), it is in one's interest to use the same random seed in the generation of the initial conditions and then ensure that the phases of the density fluctuations coincide in between the two cosmologies as this significantly reduces cosmic variance in the measurement. Specifically, if $\tilde{L}_{\rm box}$ is the comoving size of the box of the modified cosmology in units ${\rm Mpc}/\tilde{h}$ and $L_{\rm box}$ is the comoving size of the fiducial simulation box in units ${\rm Mpc}/h$, then equating the two yields $\tilde{L}_{\rm box} = L_{\rm box}\tilde{h}/h$ and ensures that the density fluctuations match on comoving scales and at all times. In other words, the comoving fundamental modes of the simulations of the modified and fiducial cosmologies are the same at all times $\tilde{k}_F = (2\pi/\tilde{L}_{\rm box})\tilde{h}/{\rm Mpc} = (2\pi/{L}_{\rm box}){h}/{\rm Mpc} = k_F$. The box size values we use are listed in Table \ref{table:params}. It is important to note that with this choice, the ratio of the spectra in the simulations of the modified and fiducial cosmologies at the same numerical value of $k$ yields actually a direct measurement of the growth-only response $G_1(k,t)$ (see e.g.~Sec.~III.~B.~of \cite{li/hu/takada} for a detailed explanation). The full isotropic response can, however, always be obtained with Eq.~(\ref{eq:R1_exp}) using the derivative of the matter power spectrum.\footnote{An alternative approach is to choose to match the {\it physical} scales, $\tilde{a}(t)\tilde{L}_{\rm box}{\rm Mpc}/\tilde{h} = a(t)L_{\rm box}{\rm Mpc}/h$, which directly measures the total response $R_1(k,t)$, but with the price that now the density fluctuations have matching scales only at a single physical time. That is, the measurement of the response at each redshift requires a set of separate universe simulations with different box sizes to be performed.}

We wish to compare the fiducial and modified cosmologies at the same physical time, but this happens at different numerical values of the scale factor, which is the coordinate that {\sc AREPO} and most $N$-body/hydro codes use to specify the output times of the particle/mass element data. Hence, if $a_{\rm out}$ represents the scale factor of the \fidu\ cosmology we are interested in, then we output the particle/mass element data in the modified cosmologies at a scale factor $a_{\rm out}^*$, defined as
\bq
t(a_{\rm out}) \equiv \int_0^{a_{\rm out}} \frac{{\rm d}a}{aH(a)} = \int_0^{a_{\rm out}^*} \frac{{\rm d}\tilde{a}}{\tilde{a}\tilde{H}(\tilde{a})}.
\eq
We will show results for $a_{\rm out} = 0.25, 0.33, 0.5, 0.66, 1.0$, which correspond to $z_{\rm out} = 3, 2, 1, 0.5, 0$. In the \high\ and \loww\ simulations, these correspond, respectively, to $a_{\rm out}^* \approx 0.249, 0.331, 0.495, 0.658,0.983$ and $a_{\rm out}^* \approx 0.251, 0.336, 0.505, 0.675, 1.017$.

Finally and important for the interpretation of our results is the fact that the hydrodynamical physics model is kept the same in between the fiducial and separate universe runs. Our results should thus be interpreted as solving nonlinear structure formation in the presence of a long-wavelength density perturbation, in a Universe with the hydrodynamical physical processes as specified by the IllustrisTNG model.\footnote{We note for completeness that the simulations make use of tabulated values of UV background radiation as a function of redshift, which should be converted to physical time for the separate universe runs. We skipped adjusting for these slight shifts in timing in the reionization history, which is slightly inconsistent, but which is not expected to have a visible impact on our total matter power spectrum measurements.}


\begin{figure*}
        \centering
        \includegraphics[width =\textwidth]{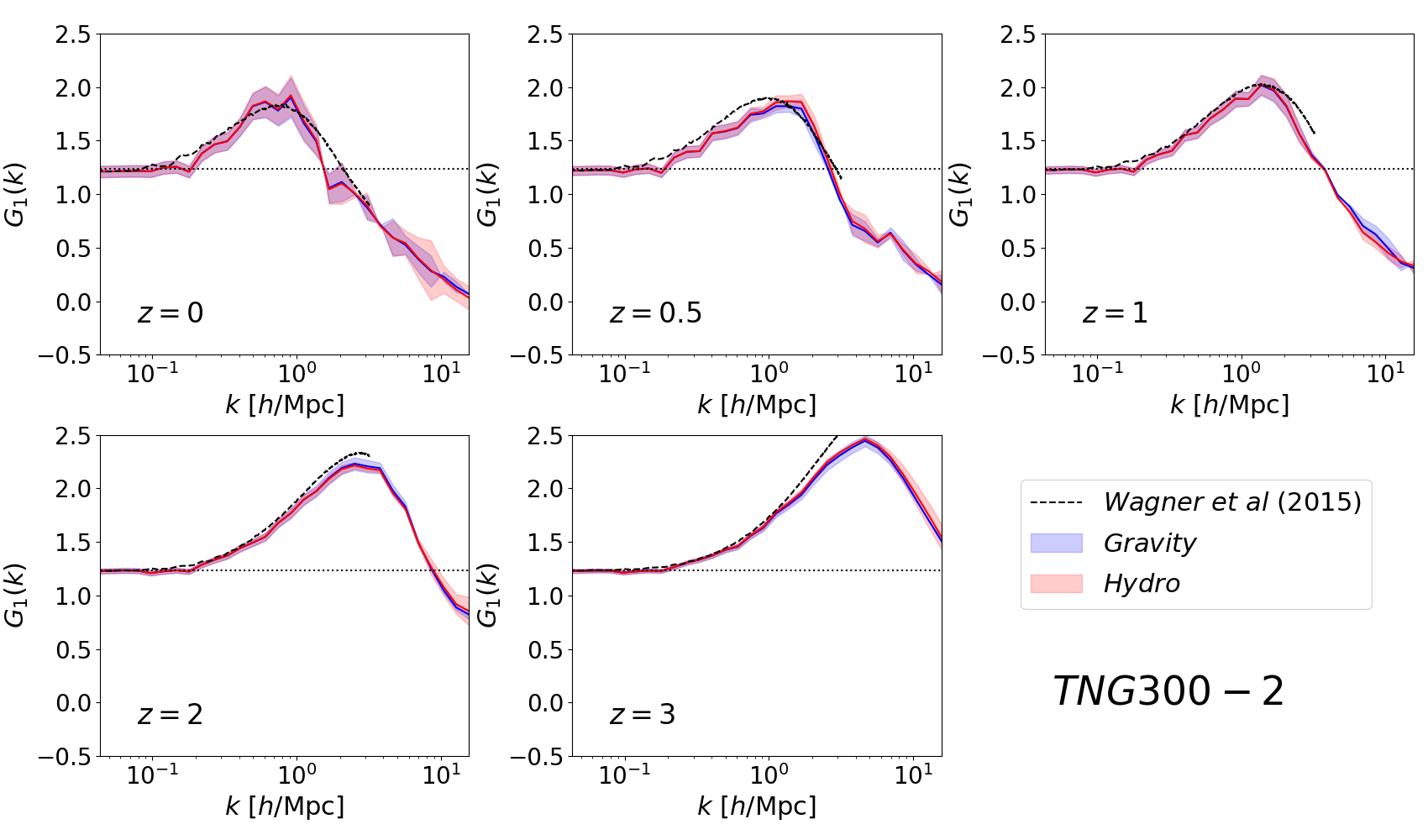}
        \caption{Growth-only first-order total matter power spectrum responses $G_1(k,z)$ measured from the Hydro and Gravity simulations of the TNG300-2 box at different redshifts, as labeled. The shaded area brackets the result obtained with the \high\ and \loww\ simulations (cf.~Eqs.~(\ref{eq:G1measure})); the corresponding solid curves mark the mean value. The black dotted line indicates the linear theory result $G_1^{\rm linear} = 26/21$. The black dashed curves show the result from gravity-only simulations for a different cosmology \citep{response}. The main takeaway is that, within the precision attained by the simulations, there is no evidence for any dependence of $G_1(k,z)$ on baryonic processes. The relative difference between the Hydro and Gravity curves is effectively always kept below $4\%$, and it does not exhibit any clear trend with scale and redshift.}
\label{fig:G1_tng300_2}
\end{figure*}

\begin{figure}
        \centering
        \includegraphics[scale=0.41]{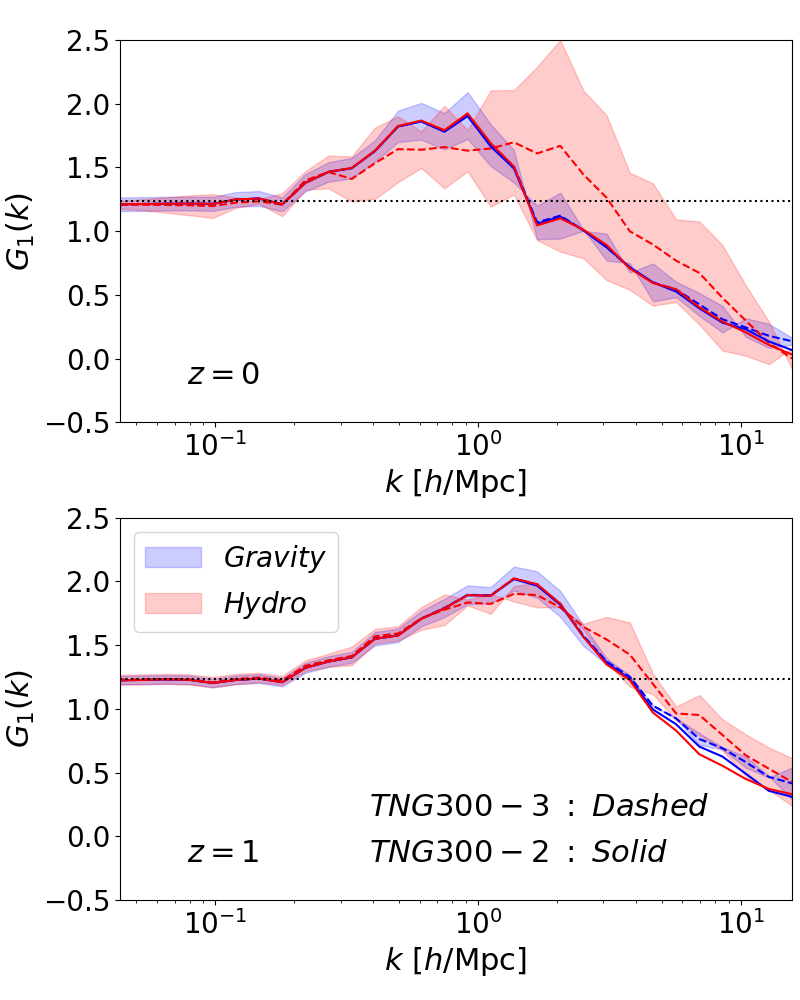}
        \caption{Same as Fig.~\ref{fig:G1_tng300_2}, but for the TNG300-3 simulation box (dashed) and fewer redshift values, as labeled. The solid lines here show the TNG300-2 result shown also in Fig.~\ref{fig:G1_tng300_2} for comparison.}
\label{fig:G1_tng300_3}
\end{figure}

\begin{figure*}
        \centering
        \includegraphics[width =\textwidth]{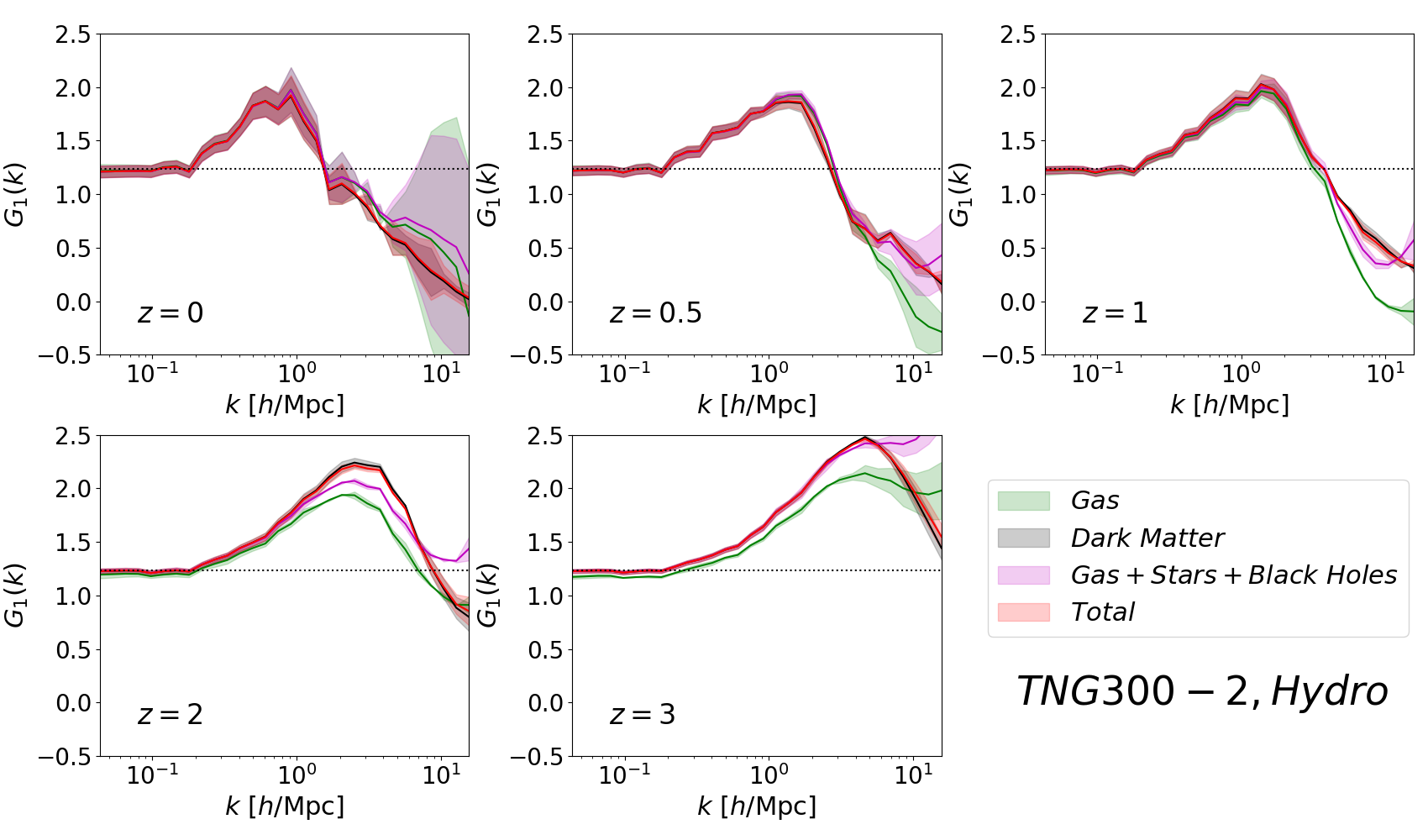}
        \caption{Same as Fig.~\ref{fig:G1_tng300_2}, but for the Hydro simulations only and for the response of the power spectrum of different mass components, as labeled (the red curves here are the same as the red curves in Fig.~\ref{fig:G1_tng300_2}).}
\label{fig:G1_tng300_2_cpts}
\end{figure*}

\section{Response measurements}\label{sec:results1}

We measure the power spectrum in the simulations with the publicly-available {\sc nbodykit} code \citep{nbodykit}. We assign the mass distribution of the particle/mass elements in the simulation onto a regular grid with $1024$ cells on a side, which is then Fourier-transformed to measure the power spectrum (for the case of a gas cell, we approximate all of its mass to lie at its geometrical center, before interpolating it to the regular grid.) In the Hydro runs, the total mass is distributed across different mass element types: gas, dark matter, stars and black holes. Unless otherwise specified, our power spectrum measurements in the Hydro runs correspond to the total mass in all mass components. We also always consider power spectra with the mass-weighted shot-noise power subtracted $P_{\rm shot} = L_{\rm box}^3 / N_{\rm eff}$, where $N_{\rm eff} = (\sum_i m_i)^2/\sum_i m_i^2$. Further, when we quote redshift values and units with $h$ retained, we always assume the \fidu\ cosmology.

Given the measured power spectrum, the first-order growth-only isotropic response function is calculated as
\bq\label{eq:G1measure}
G_1(k,z) &=& \frac{G_1^{\text{\high}}(k,z) + G_1^{\text{\loww}}(k,z)}{2},
\eq
with
\bq
\label{eq:G1measure2}G_1^{\text{\high}}(k,z) &=& \frac{1}{\delta_{L}(z)}  \left[\frac{P_m^{\text{\high}}(k,z)}{P_m^{\text{\fidu}}(k,z)} - 1\right], \\
\label{eq:G1measure3}G_1^{\text{\loww}}(k,z) &=& \frac{1}{\delta_{L}(z)}  \left[\frac{P_m^{\text{\loww}}(k,z)}{P_m^{\text{\fidu}}(k,z)} - 1\right],
\eq
where $\delta_{L}(z) = \delta_{L0}D(z)/D(z=0)$ and the numerical values of the comoving modes $k$ are the same in units of ${\rm Mpc}^{-1}$ in the \fidu, \high\ and \loww\ simulations (the superscripts indicate which cosmology each measurement corresponds to). In theory, $G_1^{\text{\high}}(k,z) = G_1^{\text{\loww}}(k,z)$, and hence, any discrepancy between the two results can be used to signal inadequacies in our numerical measurements; the difference between $G_1^{\text{\high}}$ and $G_1^{\text{\loww}}$ can also be used as a rough guide for the precision of our response measurements.

\subsection{Response of the total matter field}\label{sec:results1_total}

Figure \ref{fig:G1_tng300_2} shows $G_1(k,z)$ measured from the Hydro (red) and Gravity (blue) simulations of the TNG300-2 box. The solid lines show the result as given by Eq.~(\ref{eq:G1measure}), with the shaded area indicating the region bracketed by $G_1^{\text{\high}}$ and $G_1^{\text{\loww}}$. On large scales, both the Hydro and Gravity runs recover the expected redshift- and scale-independent linear theory result $G_1^{\rm linear}(k,z) = 26/21$. Furthermore and interestingly, the Hydro and Gravity response measurements continue to agree with one another to very good precision for all of the scales and epochs shown. This includes scales $k \gtrsim 5\ h/{\rm Mpc}$ deep in the nonlinear regime of structure formation where baryonic processes display already a marked impact on the power spectrum (cf.~Fig.~\ref{fig:boosts} below). More specifically, the relative difference between the Hydro and Gravity results is below $4\%$ and it does not exhibit any trend with redshift or scale. {There are two exceptions at $z = 0.5, k \in \left[2,4\right]\ h/{\rm Mpc}$ and $z = 1, k \in \left[5,10\right]\ h/{\rm Mpc}$, in which the difference becomes of order $6\%$ and $9\%$, respectively (not clear from the scale in Fig.~\ref{fig:G1_tng300_2}). We attribute these exceptions to statistical fluctuations associated with having only one set of phases of the initial conditions. We note also that in these exceptions, the Hydro and Gravity results remain within the shaded area of one another.} We thus report that, {\it within the precision of our measurements with the IllustrisTNG model, there is no evidence of any baryonic physics imprint on the first-order growth-only power spectrum response function.} In other words, the physical processes through which baryons alter the clustering of matter are largely insensitive to structure formation taking place at cosmic mean or in slightly overdense/underdense regions.

For reference, Fig.~\ref{fig:G1_tng300_2} shows also the result obtained by \cite{response} using gravity-only simulations with the {\sc Gadget} code \citep{2005MNRAS.364.1105S}. The small differences between their result and ours can be attributed to different cosmological parameter values. Their curve is also smoother as it is an average over 16 realizations of the initial conditions.

Figure \ref{fig:G1_tng300_3} compares the $G_1(k,z)$ measurements from the TNG300-2 (solid curves) and TNG300-3 (dashed curves and shaded areas) boxes. The results are shown for $z=0$ and $z=1$ and they illustrate that (i) the Gravity runs of both resolutions are in good agreement for all of the scales shown, but (ii) the Hydro results are somewhat discrepant on scales $k \gtrsim 2h/ {\rm Mpc}$ ($z=1$) and $k \gtrsim 0.3 h/{\rm Mpc}$ ($z=0$). The reduced accuracy of the Hydro TNG300-3 simulations in measuring the response function is not entirely surprising given its poorer performance in resolving as realistically the intricate baryonic physical processes that take place on small length scales \citep{Pillepich:2017fcc, 2018MNRAS.475..676S, Nelson:2017cxy}. The width of the red shaded area in Fig.~\ref{fig:G1_tng300_3} is also telling of this poorer resolution as it indicates that the power spectrum is responding in a qualitatively different way to an overdense and underdense perturbation, when this response should be the same. 

The lack of convergence between the Hydro runs of the TNG300-2 and TNG300-3 boxes could raise the question of whether or not the Hydro TNG300-2 results are actually converged. We can list, however, a number of reasons that convince us about the robustness of our $G_1(k,z)$ measurements at TNG300-2 resolution. First, Fig.~6 of \cite{2018MNRAS.475..676S} shows that the clustering of matter with the TNG300-2 resolution agrees to better than $2\%$ with the higher resolution simulation TNG300-1 ($N_p = 2500^3$), on the scales $k \lesssim 15 h/{\rm Mpc}$ we probe in this paper. Second, compared to the TNG300-3 case, the appreciably smaller difference between the $G_1^{\text{\high}}$ and $G_1^{\text{\loww}}$ curves (depicted by the shaded area) is also indicative of a faithful capture of the response of the power spectrum to the long-wavelength perturbation by the TNG300-2 resolution. Additionally, the nearly perfect overlap between the Gravity and Hydro TNG300-2 results shown in Fig.~\ref{fig:G1_tng300_2} (which is absent at TNG300-3 resolution) can also be invoked to strengthen further the conclusion that the result is converged. Alternatively, this would mean that any errors due to insufficient resolution were canceling almost exactly some actual physical baryonic impact on $G_1(k,z)$ on all scales and all redshifts, which we deem to be highly unlikely. Finally, the degree of numerical convergence in hydrodynamical simulations is in general redshift dependent (being better at higher redshift), but our results show no evidence of a redshift dependent difference between the $G_1(k,z)$ measured with the TNG300-2 Gravity and Hydro simulations; this is not the case at TNG300-3 resolution, for which the Gravity and Hydro results are closer at $z=1$ than at $z=0$.

\subsection{Response of individual mass components}\label{sec:results1_ind}

The results discussed thus far for the Hydro simulations correspond to the response of the total matter power spectrum computed using the total mass distribution in gas, dark matter, stars and black holes. It is also interesting, however, to measure the response of the power spectrum of the individual mass components. The result is shown in Fig.~\ref{fig:G1_tng300_2_cpts} for the gas (green) and dark matter (black) components, as well as for the mass distribution comprising gas cells, stars and black holes (magenta). The results correspond to the TNG300-2 box and the total power spectrum (red) is repeated from Fig.~\ref{fig:G1_tng300_2} for comparison. The figure shows that the dark matter power spectrum responds effectively in the same way as the total matter power spectrum; this is as expected as the total clustering of matter is dominated by the (more abundant) dark matter component. 

The response of the power spectrum of the gas component displays however a slightly different behavior. This is better noticed at redshifts $z>1$, where the additional hydrodynamical processes in which the gas participates seem to make its power spectrum respond less strongly to the presence of long-wavelength modes. These hydrodynamical phenomena include additional pressure forces felt by the gas cells, as well as the fact that the mass in gas cells is not conserved, as stars form and black holes grow in the simulations. The scales at which the gas response deviates from that of the dark matter is also redshift dependent: $k \gtrsim 3h/{\rm Mpc}\ (z=1)$, $k \gtrsim 0.8h/{\rm Mpc}\ (z=2)$, and effectively all scales shown at $z=3$, including large linear scales where the gas response plateaus at a slightly different value. Interestingly, the response of the gas+stars+black holes power spectrum is closer to that of the total power spectrum response; this is particularly noticeable at $z=3$. The mass in the gas+stars+black holes is conserved in the simulations, which suggests that the non-conservation of mass in the gas component plays an important role in the corresponding response function. A reasonable physical picture is that the enhanced star formation and black hole growth caused by the long-wavelength density perturbation leaves less mass in the gas component, effectively making its clustering less {\it responsive} to the perturbation. Once mass conservation is restored by including the mass in stars and black holes, then the corresponding response function is brought closer to that of the dark matter component (which is conserved in the TNG model). This physical picture is also in line with the fact that star formation rates peak between $z = 2$ and $z = 3$, which coincides with the epochs when the gas response differs the most from the others in Fig.~\ref{fig:G1_tng300_2_cpts}.

Regarding the responses of the gas and gas+stars+black holes components, we note that the difference between $G_1^{\rm SepHigh}$ and $G_1^{\rm SepLow}$ (shaded areas) can be non-negligible on small length scales, $k \gtrsim 4 h/{\rm Mpc}$ at $z=0$, $z=0.5$ and $z=3$, which may signal some lack of convergence of the TNG300-2 box for these components there. On larger scales, however, the width of the shaded areas is smaller (i.e., $G_1^{\rm SepHigh} \approx G_1^{\rm SepLow}$), which suggests that our results are faithful representations of the expected behavior of these response functions. A stronger case for the convergence of our gas response results at TNG300-2 resolution on scales $k \lesssim 4\ h/{\rm Mpc}$ can be made by noting that the clustering of the gas component itself is reasonably converged at this resolution. This can be seen in Fig.~5 of \cite{2018MNRAS.475..676S}, which shows that the 2-point correlation function of the gas component of the TNG300-2 resolution agrees with that of the higher resolution TNG300-1 simulation to better than $5\%$ on length scales $r\gtrsim 0.5 {\rm Mpc}/h$. Figure~A2 of \cite{Pillepich:2017jle} also shows that the gas fraction in haloes with mass $2\times10^{12} - 2\times10^{13}\ M_{\odot}/h$ found at TNG300-2 resolution (labeled TNG25-128 there) agrees to better than a few percent with that found in higher resolution runs. We note, however, that the clustering of stars and black holes at TNG300-2 resolution is not as converged as the clustering of gas and dark matter (cf.~Fig.~5 of \cite{2018MNRAS.475..676S}). Consequently, a more quantitative study of the physical picture outlined above to explain the differences between the responses of the gas and gas+stars+black holes components would certainly benefit from higher resolution response measurements.

We leave as future work a more in-depth investigation of the response of the gas component. Such a study can provide interesting theoretical insights to Lyman-alpha forest flux power spectrum analyses \citep{2006ApJS..163...80M, 2013JCAP...04..026S, 2018arXiv181203554C}, which probe the line-of-sight distribution of neutral hydrogen clouds via the Lyman-alpha absorption lines they imprint on background quasar spectra. Similarly, line intensity mapping studies \citep{2017arXiv170909066K} (including 21cm line emission \citep{2012RPPh...75h6901P, 2006PhR...433..181F, 2018ApJ...866..135V}) can also benefit from knowing how the distribution of emitting gas changes as a function of long-wavelength matter density perturbations. In fact, a number of works in these directions have taken place already with \cite{2017JCAP...06..022C} studying the response of the Lyman-alpha power spectrum to fluctuations in the quasar distribution and with \cite{2019JCAP...02..058G} investigating the response of the 21cm power spectrum signal.

\section{Baryonic effects on higher-order lensing statistics}\label{sec:results2}

The response approach to perturbation theory developed by \cite{responses1} is a formalism that uses response functions to rigorously evaluate a number of higher-order $N$-point functions in the nonlinear regime of structure formation. In this section, we use the response approach and the measurements of the baryonic impact (or lack thereof) on the power spectrum response functions discussed in Sec.~\ref{sec:results1} to predict the impact of baryonic effects on higher-order weak-lensing statistics. We will focus in particular on the case of the squeezed lensing bispectrum and the lensing super-sample power spectrum covariance. 

In the considerations below, we will limit ourselves to writing down the relevant equations without derivations; the interested reader is referred to the cited literature for details.

\subsection{Lensing equations}\label{sec:lensing}

\subsubsection{The lensing convergence power spectrum}

The lensing convergence field $\kappa$ is defined as (see e.g.~\cite{2008ARNPS..58...99H, 2015RPPh...78h6901K} for reviews on weak lensing)
\bq\label{eq:kappa}
\kappa(\vtheta) = \int_0^{\chi_S}{\rm d}\chi g(\chi) \delta_m(\vx = \chi\vtheta, z),
\eq
where $\vtheta$ is a two-dimensional position on the sky, $\chi$ is the comoving distance (we always leave implicit that $z\equiv z(\chi)$), $\delta_m$ is the three-dimensional total matter density contrast, $g(\chi) = (3/2c^2)H_0^2\Omega_m(1+z)(\chi_S-\chi)\chi/\chi_S$, $c$ is the speed of light and $\chi_S$ is the comoving distance to a single source galaxy redshift $z_S$. Under the Limber approximation (which is valid for the multipoles $\ell > 20$ we consider below), the angular power spectrum of the lensing convergence is given in terms of the three-dimensional total matter power spectrum as
\bq\label{eq:Cl}
C_\kappa(\ell) = \int_0^{\chi_S} {\rm d}\chi \frac{g(\chi)^2}{\chi^2} P_m(k_\ell, z),
\eq
where $\ell$ is the multipole moment associated with $\vtheta$ and $k_\ell = \ell/\chi$. The above equation makes apparent how the baryonic effects on the three-dimensional matter power spectrum propagate to the two-dimensional lensing one.

\subsubsection{The lensing convergence squeezed bispectrum}

The three-dimensional total matter bispectrum is defined as $B_m(\vk_1, \vk_2, \vk_3) = (2\pi)^3\delta_D(\vk_1 + \vk_2 + \vk_3) \langle\tilde{\delta}(\vk_1)\tilde{\delta}(\vk_2)\tilde{\delta}(\vk_3)\rangle_c$, where $\tilde{\delta}$ is the Fourier transform of the matter density contrast (not to be confused with the meaning of tilded quantities in Sec.~\ref{sec:theory}), $\delta_D$ is the Dirac-delta function and the subscript $_c$ indicates the correlation function is a connected one; note that the bispectrum depends in general on the relative orientation of the Fourier modes. In the so-called squeezed-limit configuration, the matter bispectrum can be written in terms of responses as
\bq\label{eq:sqB}
B_m(\vk_h, \vk_h, \vk_s; z) &=& \left[R_1(k_h,z) + \left(\mu_{\vk_h,\vk_s}^2 - \frac13\right) R_K(k_h,z) \right] \nonumber \\
&&\times P_m(k_h, z)P_m(k_s, z),
\eq
where $\vk_h$ is a small-scale (or hard) Fourier mode, $\vk_s$ is a large-scale (or soft) Fourier mode and $\mu_{\vk_h,\vk_s}$ is the cosine angle between the two wavevectors; we have also specialized to isosceles configurations for simplicity, $\vk_1 = \vk_2 = \vk_h$. Equation~(\ref{eq:sqB}) is strictly only valid for linear values of $k_s$ and $k_s \ll k_h$\footnote{More precisely, there are corrections to Eq.~(\ref{eq:sqB}) which are of order $(k_s/k_h)^2$. Note also that $P_m(k_s,z)$ is effectively the linear matter power spectrum as $k_s$ is kept in the linear regime.}, but the hard mode $k_h$ can be deep in the nonlinear regime.  Again under the Limber approximation, we can write the bispectrum of the lensing convergence in terms of the three-dimensional matter one as \citep{2001ApJ...548....7C, 2005PhRvD..72h3001D}
\bq\label{eq:Bkappa}
B_{\kappa}(\vell_1, \vell_2, \vell_3) = \int_{0}^{\chi_S} {\rm d}\chi \frac{g(\chi)^3}{\chi^4} B_m(\vk_{\ell_1}, \vk_{\ell_2}, \vk_{\ell_3}; z),
\eq
where $\vk_{\ell_i} = \vell_i/\chi$ ($i=1,2,3$). Specializing to squeezed (isosceles) configurations, $\vell_1=\vell_2=\vell_h$, $\vell_3=\vell_s$, as well as averaging over the directions of the modes $\vell_h$ and $\vell_s$ permits one to write
\bq\label{eq:Bkappa_2}
B_\kappa(\ell_h, \ell_h, \ell_s) &=& \int_0^{2\pi} \int_0^{2\pi}\frac{{\rm d}\varphi_{\vell_h}}{2\pi} \frac{{\rm d}\varphi_{\vell_s}}{2\pi} B_{\kappa}(\vell_h, \vell_h, \vell_s) \nonumber \\
&=& \int_{0}^{\chi_S} {\rm d}\chi \frac{g(\chi)^3}{\chi^4} R_{\perp}(k_h, z) P_m(k_h, z)P_m(k_s, z), \nonumber \\
\eq
where $R_{\perp} = R_1 + R_K/6$ and the equation is valid up to corrections that scale as $\left(\ell_s/\ell_h\right)^2$. This equation shows that the impact of baryonic effects on the squeezed-lensing bispectrum is controlled by the impact of baryonic effects on the small-scale total matter power spectrum $P_m$, as well as on the responses $R_1$ and $R_K$. In Sec.~\ref{sec:results1}, we have seen that the growth-only response $G_1(k,z)$ is effectively independent of baryonic effects, which implies that the baryonic effects on $R_1$ are specified by the modifications to $P_m$ via the logarithmic derivative term ${\rm dln}P_m(k,z)/{\rm dln}k$ in Eq.~(\ref{eq:R1_exp}). Regarding the tidal response $R_K$ of Eq.~(\ref{eq:RK_exp}), we have not explicitly checked whether the corresponding growth-only response $G_K$ depends on baryonic effects, but the lack of any impact on $G_1$ suggests any such modification on $G_K$ should be negligible as well.\footnote{The tidal response $R_K$ can also be evaluated with separate universe methods, but this requires modifying the N-body/hydro codes to allow them to follow the anisotropic expansion of the local spacetime induced by the long-wavelength tidal field; see \cite{andreas} for the details of such an implementation in gravity-only simulations.} We proceed with this assumption on $G_K$, in which case the way $R_K$ depends on baryonic effects is specified by the power spectrum according to Eq.~(\ref{eq:RK_exp}). Consequently, the whole dependence of the squeezed-lensing bispectrum on baryonic effects can be predicted using the same dependence on the power spectrum.

\subsubsection{The lensing covariance matrix}

Another interesting and powerful application of the response approach is in the calculation of the covariance matrix of the lensing convergence power spectrum \citep{completessc, nocng}. For a given power spectrum estimator $\hat{C}(\ell)$, the covariance is defined as $\cov_\kappa(\ell_1, \ell_2) = \langle\hat{C}(\ell_1)\hat{C}(\ell_2)\rangle - \langle\hat{C}(\ell_1)\rangle\langle\hat{C}(\ell_2)\rangle$, and it can be split into three physically distinct contributions known as the Gaussian term (G), the super-sample covariance term (SSC) and the connected non-Gaussian term (cNG):
\bq\label{eq:cov_dec}
\cov_\kappa(\ell_1, \ell_2) = \cov_\kappa^{G}(\ell_1, \ell_2) + \cov_\kappa^{SSC}(\ell_1, \ell_2) + \cov_\kappa^{cNG}(\ell_1, \ell_2). \nonumber \\
\eq
The covariance matrix is a crucial ingredient in parameter inference analyses using weak lensing data, which makes it observationally relevant to ask how baryonic effects on the covariance may affect parameter constraints. 

Ignoring a number of real-life complications, such as masking or non-uniform survey depth, the Gaussian term can be written as
\bq\label{eq:g}
\cov_\kappa^{G}(\ell_1, \ell_2) = \frac{4\pi}{\Omega_W\ell_1\Delta \ell_1} \left[C(\ell_1) + C_{\rm shape}\right]^2\delta_{ij},
\eq
where $\Delta \ell_1$ is the size of the multipole bin in which the power spectrum is measured, $\delta_{ij}$ is a Kronecker delta that ensures the result is only non-zero if the two modes $\ell_1, \ell_2$ are in the same bin, $\Omega_W = 4\pi f_{\rm sky}$ with $f_{\rm sky}$ the surveyed sky fraction and $C_{\rm shape}$ is a shot noise component that arises due to the unknown (i.e. unlensed) shapes of the source galaxies. A key point to note is that the impact of baryonic effects on the Gaussian term is uniquely determined by the way $C(\ell_1)$ depends on baryonic effects, which we can readily predict using Eq.~(\ref{eq:Cl}). When we evaluate this term below, we do so for 50 $\ell$ bins equally distributed in log-scale between $\ell_{\rm min}=20$ and $\ell_{\rm max}=5000$, $f_{\rm sky} = 0.363$ and $C_{\rm shape} = \sigma_e^2/2/\bar{n}_{\rm gal}$, with a RMS source galaxy ellipticity of $\sigma_e = 0.37$ and $\bar{n}_{\rm gal} = 30/{\rm arcmin}^2$ (these values are inspired by the expected specifications for surveys like Euclid).

Under the Limber approximation, the lensing SSC term can be written as \citep{takada/hu:2013, completessc}
\bq\label{eq:ssc}
\cov_\kappa^{SSC}(\ell_1, \ell_2) &=& \frac{1}{\Omega_W^2} \int_0^{\chi_S} {\rm d}\chi \frac{g(\chi)^4}{\chi^6} \int \frac{{\rm d}^2\vell}{(2\pi)^2} |\tilde{W}(\vell)|^2 P_m(k_{\ell}) \nonumber \\
&\times& R_{\perp}(k_{\ell_1},z) R_{\perp}(k_{\ell_2},z) P_m(k_{\ell_1},z)P_m(k_{\ell_2},z), \nonumber \\
\eq
where $\tilde{W}(\vell)$ is the Fourier transform of the survey footprint on the sky.\footnote{\cite{completessc} has actually shown that the Limber approximation can underestimate the amplitude of the small-scale SSC term by $\approx 10\%$ for $f_{\rm sky} \approx 0.3$. Here, we opt to display and evaluate the equations in the Limber approximation because they are slightly simpler and because they depend on the exact same physical ingredients. Our conclusions on the corresponding impact of baryonic effects thus hold for both derivations.} This term is part of the connected four-point function that defines the covariance matrix of a two-point function and its physical meaning is that it describes the coupling between the nonlinear sub-survey modes, $\ell_1, \ell_2$, with linear super-survey modes $\vell$ that are integrated over. Similarly to the squeezed lensing bispectrum case, knowing how the power spectrum and its first-order responses $R_1$ and $R_k$ depend on baryonic effects permits one to straightforwardly evaluate how the SSC term is modified by the same effects. In the numerical results below we consider a disk-shaped survey window function for which $|\tilde{W}(\vell)|^2 = \Omega_W^2\left[2J_1(\ell\theta_W)/(\ell\theta_W)\right]^2$, where $\theta_W = \sqrt{\Omega_W/\pi}$ and $J_1$ is the first-order spherical Bessel function.

Finally, the cNG term corresponds to the rest of the connected four-point function that does not belong to the SSC term; effectively, it describes the coupling between the sub-survey modes $\ell_1, \ell_2$ that arises as the density field becomes less Gaussian distributed. \cite{nocng} showed recently that for the survey specifications of future (as well as current) surveys, this term is likely to play a negligible role in cosmological analysis using weak-lensing data, and for that reason, we skip evaluating it numerically in this paper. We note for completeness that the majority of this term can be evaluated with second-order power spectrum response functions \citep{responses2}.

\begin{figure}
        \centering
        \includegraphics[scale = 0.41]{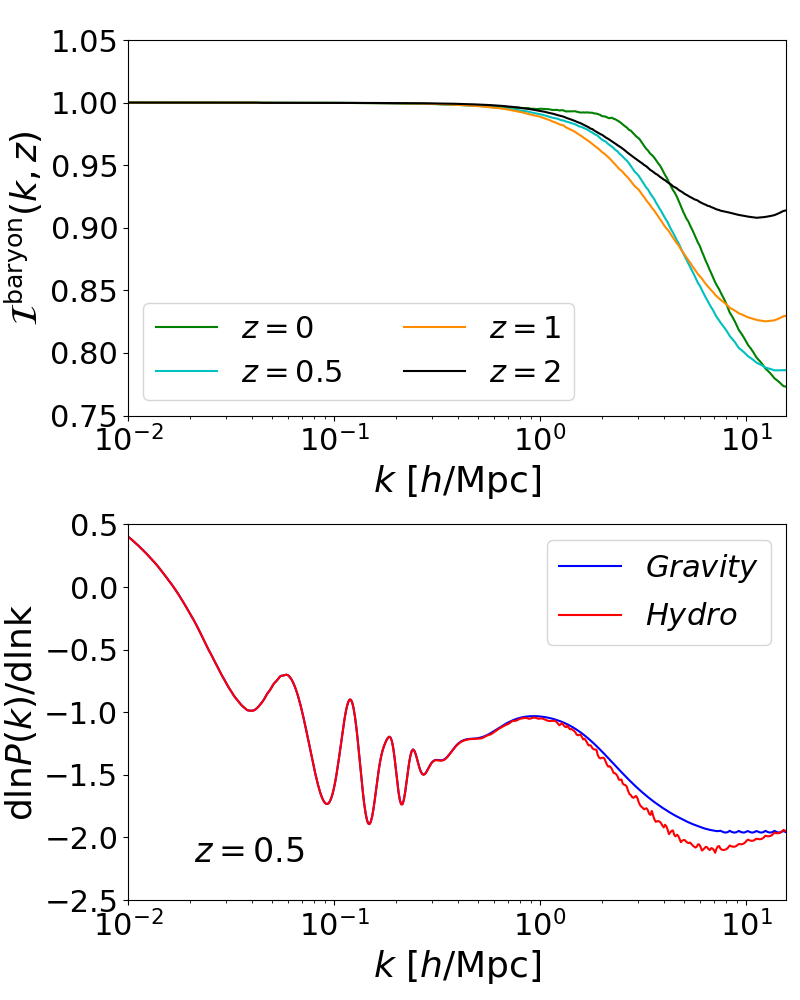}
        \caption{The upper panel shows the {\it baryon boost factor} defined in Eq.~(\ref{eq:bbf}) as the ratio of the power spectrum of the Hydro and Gravity runs of the TNG300-2 simulations. The curves are for different redshifts, as labeled. The lower panel shows the logarithmic derivative term ${\rm dln}P_m(k,z)/{\rm dln}k$ that enters the first-order density $R_1(k,z)$ and tidal $R_K(k,z)$ response functions (cf.~Eqs.~(\ref{eq:R1_exp}) and (\ref{eq:RK_exp})); the result is shown at $z=0.5$ and for cases with and without baryonic effects, as labeled.}
\label{fig:boosts}
\end{figure}

\begin{figure}
        \centering
        \includegraphics[scale=0.41]{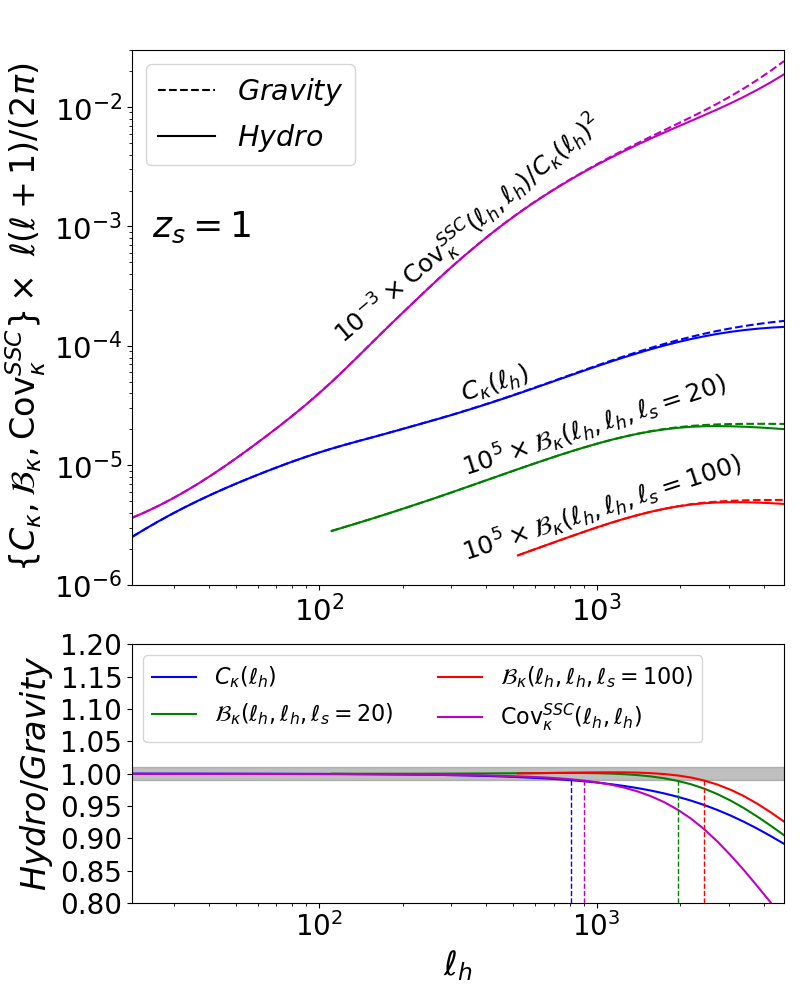}
        \caption{Impact of baryonic effects on the lensing power spectrum (blue), squeezed-lensing bispectrum (for two values of the soft wavenumber, red and green) and diagonal of the lensing SSC term (magenta), as labeled. The solid and dashed lines show the results with and without baryonic effects taken into account, respectively. The curves in the upper panel have been rescaled for visualization purposes (when displaying the SSC term, we use the same power spectrum in the ratio). The lower panel shows the ratio of the Hydro and Gravity cases. The vertical dashed lines indicate when the Hydro and Gravity cases deviate by more than $1\%$ (same color coding). The squeezed bispectrum curves are only plotted for $\ell_h > 5\ell_s$ to ensure the configuration is sufficiently squeezed (cf.~Eq.~(\ref{eq:Bkappa_2})).}
\label{fig:lensing}
\end{figure}

\begin{figure*}
        \centering
        \includegraphics[scale=0.41]{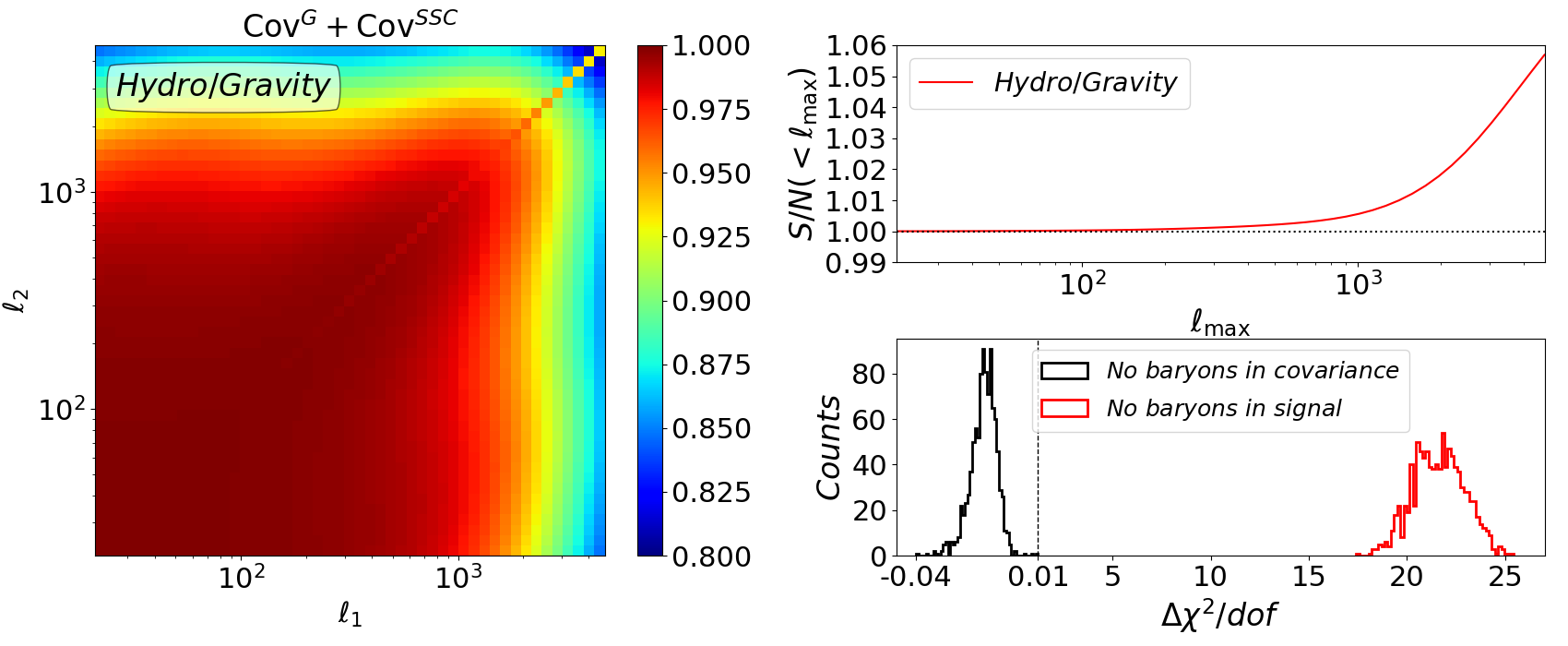}
        \caption{The left panel shows the ratio of the (G+SSC) lensing covariance matrix evaluated with baryonic effects to that evaluated without baryonic effects. The upper right panel shows the ratio of the cumulative signal-to-noise (cf.~Eq.~(\ref{eq:sn})) with baryonic effects taken into account to that without baryonic effects. The lower right panel shows the impact on the goodness-of-fit $\chi^2$ (cf.~Eq.~(\ref{eq:chi2})) from ignoring baryonic effects on the calculation of the signal lensing power spectrum (red) and on the calculation of the lensing covariance matrix (black). The number of degrees of freedom is the number of multipole bins $dof = 50$. Note that the distributions are shown in two linear scales (separated by the vertical dotted line) to facilitate visualization.}
\label{fig:covariance}
\end{figure*}

\subsection{Numerical results}\label{eq:numerical}

\subsubsection{Baryonic effects on higher-order lensing statistics}

As mentioned above, the dependence of the squeezed-lensing bispectrum and of the lensing SSC term on baryonic effects is encapsulated by the corresponding dependence of the power spectrum. This happens via the power spectrum $P_m(k,z)$ itself that enters the above equations, as well as via the logarithmic derivative ${\rm dln}P_m(k,z)/{\rm dln}k$ that enters $R_1(k,z)$ and $R_K(k,z)$ (cf.~Eqs.~(\ref{eq:R1_exp}) and (\ref{eq:RK_exp})). Evaluating the latter derivative numerically using the power spectrum measured from a simulation would however yield too noisy a result. To circumvent this, in our numerical results, we evaluate power spectra as
\bq\label{eq:Pm_eval}
P_m(k,z) = P_m^{\rm Emu.}(k,z) \mathcal{I}^{\rm baryon}(k,z),
\eq
where $P_m^{\rm Emu.}(k,z)$ is the gravity-only power spectrum evaluated using the {\sc CosmicEmu} emulator \citep{emulator, emulator2} and $\mathcal{I}^{\rm baryon}(k,z)$ is a {\it baryon boost factor} defined as the ratio of the power spectrum from the IllustrisTNG Hydro and Gravity runs
\bq\label{eq:bbf}
\mathcal{I}^{\rm baryon}(k,z) = \frac{P_m^{\rm TNG,Hydro}(k,z)}{P_m^{\rm TNG,Gravity}(k,z)}.
\eq
This ratio is appreciably smoother than the power spectra measurements themselves, which together with the also smooth emulator curves yields sufficiently noise-free derivatives ${\rm dln}P_m(k,z)/{\rm dln}k$. We use the TNG300-2 results to evaluate $\mathcal{I}^{\rm baryon}(k,z)$ at the available redshifts, from which we interpolate to carry out the line-of-sight integrals that characterize the calculation of the lensing quantities.  We have checked that the {\sc CosmicEmu} and TNG300-2 gravity-only spectra agree to better than $4\%$ at $z=0, 1$ on scales $k < 5 h/{\rm Mpc}$. We note, however, that the accuracy of the gravity-only spectra is not critical for the interpretation of our results, which focus on the relative impact of baryonic effects and not on absolute spectra values.

The scale-dependence of $\mathcal{I}^{\rm baryon}(k,z)$ is shown in the upper panel of Fig.~\ref{fig:boosts} for $z=0, 0.5, 1, 2$, as labeled. The result shown displays the well known suppression effect due to AGN feedback that becomes important on scales typically inside dark matter haloes $k \gtrsim 1 h/{\rm Mpc}$. For instance, on scales $k = 3-5 h/{\rm Mpc}$, the baryonic effects in the TNG300-2 simulations induce a suppression of power that is of order $10-15\%$ at $z=0.5, 1$. Note also the non-trivial interplay between the scale- and redshift-dependence of $\mathcal{I}^{\rm baryon}(k,z)$; for instance, at $k\approx10 h/{\rm Mpc}$ there is a stronger effect at $z=0$ than $z=2$, but the situation is reversed at $k\approx2-3 h/{\rm Mpc}$. These quantitative results are in line with those reported in \cite{2018MNRAS.475..676S}. The lower panel of Fig.~\ref{fig:boosts} shows the corresponding impact of the baryonic effects on ${\rm dln}P_m(k,z)/{\rm dln}k$ ($z=0.5$), which is the quantity that enters explicitly the first-order responses $R_1(k,z)$ and $R_K(k,z)$ in Eqs.~(\ref{eq:R1_exp}) and (\ref{eq:RK_exp}), respectively.

Figure \ref{fig:lensing} shows the imprint that baryonic effects cast on the lensing power spectrum $C_\kappa(\ell_h)$ (blue), the squeezed-lensing bispectrum $B_\kappa(\ell_h, \ell_h, \ell_s)$ (red and green for two values of $\ell_s$) and the diagonal of the SSC term $\cov_\kappa^{SSC}(\ell_h, \ell_h)$, for a single source redshift $z_S=1$, as labeled. The figure shows that the suppression effects on the small-scale three-dimensional power spectrum shown in Fig.~\ref{fig:boosts} manifest themselves also in the line-of-sight-projected lensing quantities shown in Fig.~\ref{fig:lensing}. There are some differences in the exact size and scale-dependence of the effect, whose origin can be traced back to two main factors. One is the fact that the line-of-sight integrands of $C_{\kappa}$, $B_{\kappa}$ and $\cov^{SSC}_{\kappa}$ peak at different redshifts, and hence, are affected differently by the time-dependent baryonic effects on $P_m(k,z)$; for example, the integrands of the SSC term tend to peak at lower redshifts (not shown). The other is that the calculation of $B_{\kappa}$ and $\cov^{SSC}_{\kappa}$ depends also on the responses $R_1(k,z)$ and $R_K(k,z)$, which are actually enhanced (not suppressed) by the baryonic effects. This is because baryons make the logarithmic derivative ${\rm dln}P_m(k,z)/{\rm dln}k$ more negative (cf.~Fig.~\ref{fig:boosts}), but this derivative enters $R_1(k,z)$ and $R_K(k,z)$ with a negative sign (cf.~Eqs.~(\ref{eq:R1_exp}) and (\ref{eq:RK_exp})). Quantitatively, for our idealized lensing setup, Fig.~\ref{fig:lensing} shows that the SSC term is that which is most strongly affected by the baryonic effects, whose size exceeds $1\%$ ($5\%$) on multipoles $\gtrsim 900$ ($\gtrsim 2000$). The squeezed-bispectrum is that which is the least affected with the same effects exceeding $1\%$ ($5\%$) on multipoles $\gtrsim 2000$ ($\gtrsim 4000$). At a fixed $\ell = 3000$, the suppression induced by baryons on the lensing power spectrum, squeezed-bispectrum and SSC term is at the $6\%$, $2-4\%$ and $12\%$ levels, respectively.

We have explicitly checked (although not shown) that these quantitative results for $z_S=1$ do not change critically if we choose other source redshift values. For higher $z_S$, the impact of baryons gets reduced because the lensing integrands peak at higher redshift, where the impact of baryons on the three-dimensional power spectrum is generically weaker (cf.~Fig.~\ref{fig:boosts}). This effect is however not very large because of the broad lensing integrands. For instance, the scales above which the impact on the SSC term exceeds $1\%$ are $\ell \approx 500, 900, 1100$, for $z_S=0.5, 1, 2$, respectively; for the lensing squeezed-bispectrum, the same values are $\ell \approx 1100, 2000, 2500$. Further, the fact that the lensing SSC term is more strongly affected than the power spectrum, and that the latter is in turn more strongly affected than the lensing squeezed-bispectrum holds equally for $z_S=0.5, 1, 2$. The behavior of the baryonic effects within $z_S \in \left[0.5, 2\right]$ should be indicative of what is expected from going beyond the idealized single source redshift scenario we adopted here to the redshift distributions of current and future surveys (which span roughly this redshift range).

It is interesting to compare our squeezed lensing bispectrum results to those of \cite{2013MNRAS.434..148S}, who investigated the impact of baryonic effects on lensing three-point statistics as well. They work with the third-order $\langle M^3_{\rm ap}(\theta)\rangle$ aperture mass statistic \citep{2003ApJ...592..664P, 2004MNRAS.352..338J, 2005A&A...431....9S}, which is sensitive to general configurations of the lensing bispectrum (not just squeezed, as the case here using the response approach).\footnote{See \cite{Semboloni:2010er, 2013MNRAS.430.2476S, 2014MNRAS.441.2725F} for a number of observational analyses of $\langle M^3_{\rm ap}(\theta)\rangle$, none of which explicitly take baryonic effects into account however.} They evaluate the three-dimensional bispectrum using an approximate model based on perturbation theory and explore a number of different implementations of the baryonic physical processes present in the {\sc OWLS} simulations. They also consider more realistic galaxy source distributions, compared to our idealized single source redshift scenario. \cite{2013MNRAS.434..148S} reported that the baryonic effects can have an impact on the three-point statistic that may be up to $\approx2$ times larger than that on the two-point statistic (see e.g.~Fig.~1 in \cite{2013MNRAS.434..148S}), and that this distinct sensitivity could be useful in the design of baryon mitigation strategies in real data analyses. Our results in Fig.~\ref{fig:lensing} display, on the other hand, a weaker baryonic impact on $B_\kappa(\ell_h, \ell_h, \ell_s)$ than on $C_\kappa(\ell_h)$. We note however that the differences in methodology mentioned above between the two works make it hard to establish a robust comparison.

\subsubsection{Baryonic effects on parameter inference using weak-lensing data}

The result shown in Fig.~\ref{fig:covariance} is useful to understand the impact that baryonic effects can have on parameter inference analysis using weak-lensing data. The left panel shows the G+SSC lensing covariance matrix (cf.~Eq.~(\ref{eq:cov_dec})) for the idealized lensing setup described in Sec.~\ref{sec:lensing}; specifically, it shows the ratio of the matrix calculated with baryonic effects to that without baryonic effects taken into account. The upper right panel shows the impact that baryonic effects on the covariance calculation leave on the cumulative signal-to-noise ratio
\bq\label{eq:sn}
\left(\frac{S}{N}\right)^2(\ell_{\rm max}) = \sum_{\ell_1, \ell_2 < \ell_{\rm max}} C_{\kappa}(\ell_1) \cov_\kappa^{-1}(\ell_1, \ell_2) C_{\kappa}(\ell_2),
\eq
for fixed $C_{\kappa}$ calculation with baryonic effects included. Figure~\ref{fig:covariance} shows that taking baryonic effects into account lowers the amplitude of the covariance matrix entries on small-scales, which effectively increases the signal-to-noise of the analysis. For the case of our idealized single source redshift lensing setup, the increase is modest, $\approx 5\%$ for $\ell_{\rm max} = 5000$, but an interesting way to interpret this result is that ignoring baryonic effects on the weak-lensing covariance calculation represents an overestimation of the total statistical error budget, and it is thus conservative from the point of view of inferred parameter error bars. 

In addition to controlling the size of the error bars on parameters, a miscalculation of the covariance matrix can in general also bias the central value of the parameter constraints. To estimate this effect we have drawn $1000$ realizations of a convergence power spectrum data vector $C^{{\rm data}}_{\kappa}(\ell_i)$ (where $i$ runs over the $50$ multipole bins) from a multivariate Gaussian distribution with a mean $C_{\kappa}(\ell)$ and $\cov_{\kappa}(\ell_1, \ell_2)$ (G+SSC) evaluated with baryonic effects taken into account. For each sampled data vector we have evaluated the $\chi^2$ quantity
\bq\label{eq:chi2}
\chi^2 = \sum_{i,j=1}^{50}\left(C_\kappa(\ell_i) - C_\kappa^{\rm data}(\ell_i)\right) \cov_{\kappa}^{-1}(\ell_i, \ell_j) \left(C_\kappa(\ell_j) - C_\kappa^{\rm data}(\ell_j)\right) \nonumber \\
\eq
for three distinct cases:
\begin{enumerate}
\item {\it Baryons in signal and covariance}: both $C_\kappa$ and $\cov_{\kappa}$ are calculated with baryonic effects taken into account;
\item {\it No baryons in the covariance}: the calculation of $C_\kappa$ takes baryonic effects into account, but the calculation of $\cov_{\kappa}$ does not;
\item {\it No baryons in the signal}: the calculation of $C_\kappa$ does not take baryonic effects into account, but the calculation of $\cov_{\kappa}$ does.
\end{enumerate}
The lower right panel of Fig.~\ref{fig:covariance} shows the histogram of the difference between the $\chi^2$ of cases (ii) and (iii) above to that of case (i). Ignoring the impact that baryonic effects can have on the signal (red) induces an appreciable degradation of the goodness-of-fit $\Delta\chi^2/{dof} \approx 22$. In real data analysis, this would trigger shifts in the cosmological parameter values away from their {\it true} values to attempt minimizing the $\chi^2$ value. This well-known result is what justifies the many efforts currently being undertaken to incorporate baryonic physics in parameter inference analysis using weak-lensing spectra  \citep{Semboloni:2011fe, 2015JCAP...12..049S, 2015MNRAS.454.2451E, 2015MNRAS.454.1958M, 2015MNRAS.450.1212H, 2018MNRAS.480.3962C, 2018arXiv180901146H, 2018arXiv181008629S}. On the other hand, neglecting baryonic effects on the covariance matrix results in an appreciably smaller change in the goodness-of-fit $\Delta\chi^2/{dof} \approx -0.02$. This is depicted by the distribution shown in black in the lower right panel of Fig.~\ref{fig:covariance}, which is plotted with a different x-axis scale to permit visualizing the shape of the distribution. 

We note that these considerations, although indicative of a small practical impact of baryonic effects in weak-lensing covariance matrices, refer to our idealized lensing setup. More robust, quantitative and survey-specific conclusions should be drawn from more realistic simulated likelihood analyses including lensing tomography, realistic source galaxy distributions and systematic effects (intrinsic alignments, photo-$z$, etc.), as well as measuring the impact of baryonic effects at the level of parameter constraints. The fact that baryonic effects,  in particular AGN feedback, are expected to lower the small-scale entries of lensing covariance matrices is however robust to the exact choice of the source redshift distribution.

\section{Summary and Conclusions}\label{sec:sum}

In this paper, we have used the separate universe simulation technique applied to the IllustrisTNG hydrodynamical galaxy formation model to measure the impact of baryonic physical processes on matter power spectrum response functions. The responses specify the scale- and redshift-dependence of the {\it response} of the small-scale matter power spectrum to the presence of long-wavelength density and tidal field fluctuations. The separate universe formalism is a technique that incorporates these long-wavelength modes via a redefinition of the cosmology that is actually simulated (cf.~Sec.~\ref{sec:theory_sepuni}). 

We focused on measurements of the first-order isotropic growth-only power spectrum response function $G_1(k,z)$, which together with the matter power spectrum $P_m(k,z)$, fully specifies the first-order isotropic response $R_1(k,z)$ (cf.~Eq.~(\ref{eq:R1_exp})). Armed with the knowledge of the baryonic effects on $P_m(k,z)$, $R_1(k,z)$, as well as on the first-order tidal response $R_K(k,z)$, one can then use the machinery of the response approach to perturbation theory to straightforwardly evaluate how baryonic effects  impact a number of higher-order matter/lensing correlation functions in the nonlinear regime of structure formation (cf.~Sec.~\ref{sec:lensing}). 

We have carried out separate universe simulations on the TNG300 box ($L_{\rm box} = 205h/{\rm Mpc}\approx 300{\rm Mpc}$) at two particle/mass element resolutions: $N_p=1250^3$ (TNG300-2) and $N_p = 625^3$ (TNG300-3). For each, we have run hydrodynamical simulations with the full TNG physics model (which we dubbed Hydro runs), as well as gravity-only counterparts (which we dubbed Gravity runs). While the cosmological parameters of the simulation are modified to mimic the presence of long-wavelength density perturbations, the remaining parameters of the TNG galaxy formation model are held fixed. Our measurements describe therefore the response of the matter power spectrum at fixed galaxy formation physics model.

Our main results can be summarized as follows:
\begin{itemize}
\item The growth-only total matter power spectrum response $G_1(k,z)$ is independent of baryonic effects on redshifts $z<3$ and scales $k \lesssim 15 h/{\rm Mpc}$ (cf.~Fig.~\ref{fig:G1_tng300_2}). This means that the dependence of the full isotropic response $R_1(k,z)$ on baryonic effects is totally encoded in the term $\propto {\rm dln}P_m(k,z)/{\rm dln}k$ (cf.~Fig.~\ref{fig:boosts} and Eq.~(\ref{eq:R1_exp})).

\item The $G_1(k,z)$ response of the power spectrum of the dark matter component alone is nearly indistinguishable of the response of the total matter fluid (cf.~Fig.~\ref{fig:G1_tng300_2_cpts}). Our results suggest, however, that the hydrodynamical physical processes felt by the gas cells (e.g.~loss of mass to form stars or fuel black hole growth) lower the amplitude of its $G_1(k,z)$ response w.r.t.~those of the total and dark matter components. These differences are also more pronounced at higher redshift (cf.~Fig.~\ref{fig:G1_tng300_2_cpts}).

\item The exact size and scale-dependence of the impact of baryonic physics on the lensing squeezed-bispectrum and super-sample covariance term can differ from that of the power spectrum (cf.~Fig.~\ref{fig:lensing}). For instance, for a single source redshift of $z_S=1$ and at $\ell = 3000$, baryonic effects suppress the amplitude of the lensing power spectrum by $\approx 6\%$, the lensing squeezed-bispectrum by $\approx 2-4\%$ and the diagonal of the SSC term by $\approx 12\%$.

\item Skipping the incorporation of baryonic effects in the calculation of weak-lensing covariance matrices overestimates its amplitude on small scales, and is thus conservative from the point of view of inferred parameter error bars. In other words, gravity-only covariances yield artificially larger parameter errors, even if only by small amounts (cf.~Fig.~\ref{fig:covariance}). Further, ignoring baryonic effects on the covariance has a negligible impact on the $\chi^2$ goodness-of-fit, compared to the impact from ignoring the same effects on the modeling of the signal.
\end{itemize}

The steps that we have taken in this paper to obtain our results can prove useful in the design of strategies to incorporate or mitigate the impact of baryonic physics in real data analyses. This is the case, for instance, at the level of the modeling of the signal in cosmological inferences using the lensing bispectrum, as well as at the level of current observational data analyses using the lensing power spectrum in the evaluation of its covariance matrix. A noteworthy advantage of using the response approach to perturbation theory is that it permits one to study higher-order statistics in the nonlinear regime with nearly the same numerical costs as two-point function studies.

Finally, beyond the study of first-order matter power spectrum responses, there are a number of additional interesting investigations that can be carried out with separate universe simulations of the IllustrisTNG model. An example is the exploration of larger values of the amplitude $\delta_{L0}$ of the long-wavelength mode to measure higher-order power spectrum response functions; these find applications in the calculation of the cNG term of the matter power spectrum covariance \citep{responses2} and matter bispectrum covariance \citep{sqbcov}. Another example is the study of galaxy bias as the response of the galaxy number density to the presence of long-wavelength modes \citep{lazeyras/etal, li/hu/takada:2016, baldauf/etal:2015}; this would be complementary to the galaxy bias measurements performed already by \cite{2018MNRAS.475..676S} using the large-scale limit of galaxy clustering. Similarly, the study of galaxy power spectrum responses with realistic galaxy formation processes becomes also possible and it can provide interesting insights on galaxy power spectrum covariances, as well as on the squeezed galaxy bispectrum that is a relevant observable in studies of primordial non-Gaussianity.

\section*{Acknowledgements}

We would like to thank An\u{z}e Slosar for useful comments and discussions. The simulations used in this work were run on the Cobra supercomputer at the Max Planck Computing and Data Facility (MPCDF) in Garching near Munich. AB and FS acknowledge support from the Starting Grant (ERC-2015-STG 678652) "GrInflaGal" of the European Research Council.


\bibliographystyle{mn2e}
\bibliography{REFS}


\bsp	
\label{lastpage}
\end{document}